\documentstyle[12pt,preprint]{aastex}

\slugcomment{Submitted to ApJ Sep 03, 2001; Accepted Nov 16}

\lefthead{}
\righthead{}

\begin{document}

\title{Detection of HI 21 cm-line absorption in the Warm Neutral \\ 
       Medium and in the Outer Arm of the Galaxy}

\author{K. S. Dwarakanath}
\affil{Raman Research Institute, Bangalore 560 080, India}
\email{dwaraka@rri.res.in}

\and

\author{C. L. Carilli and W. M. Goss}
\affil{National Radio Astronomy Observatory, PO Box O, Socorro, NM 87801}
\email{ccarilli@aoc.nrao.edu, mgoss@aoc.nrao.edu}

\shorttitle{HI absorption in WNM and Outer Arm}
\shortauthors{Dwarakanath, Carilli, and Goss}

\begin{abstract}

Using the Westerbork Synthesis Radio Telescope, we have detected HI 21 cm-line
absorption in the Warm Neutral Medium of the Galaxy toward the extragalactic source
3C147. This absorption, at an LSR velocity of --29$\pm$4 km s$^{-1}$ with a 
full width at half maximum of 53$\pm$6 km s$^{-1}$, is associated with 
the Perseus Arm of the Galaxy. 
The observed optical depth is (1.9$\pm$0.2)$\times$10$^{-3}$. The estimated spin temperature
of the gas is 3600$\pm$360 K. The volume density is 0.4 cm$^{-3}$ assuming
pressure equilibrium. Toward two other sources,  3C273 and 3C295, no
wide HI 21 cm-line absorption was detected. The highest of the 3$\sigma$ 
lower limits on the spin 
temperature of the Warm Neutral Medium is 2600 K. In addition, we have also detected HI
21 cm-line absorption from high velocity clouds in the Outer Arm toward 3C147 and
3C380 at LSR velocities of --117.3, --124.5 and --113.7 km s$^{-1}$ respectively. 
We find two distinct temperature components in the high velocity clouds with spin
temperatures of greater than 1000 K and less than 200 K, respectively. 

\end{abstract}


\section{INTRODUCTION}
The early HI 21 cm-line absorption and emission studies of the Galaxy lead to the 
two-phase model of the interstellar medium in which dense
cool HI clouds (the Cold Neutral Medium) were in pressure equilibrium 
with a warm low-density medium (the Warm Neutral Medium (WNM))
(Clark, Radhakrishnan, \& Wilson 1962, Clark 1965, Radhakrishnan et al. 1972).
The mean spin temperature of the cold clouds was estimated to be 80 K.
Only a lower limit to the spin temperature of the WNM of 1000 K was
obtained from an upper limit to the optical depth of HI 21 cm-line absorption in
the WNM. Since then there have been several attempts to measure the optical depth 
of the WNM (Mebold \& Hills 1975, Kalberla, Mebold, \& Reich 1980, Liszt, Dickey,
\& Greisen 1982,
Mebold et al. 1982, Payne, Salpeter, \& Terzian 1983, Kulkarni et al. 1985). 
Using the Bonn Telescope, 
Mebold \& Hills (1975) estimated an optical depth for the WNM based on the
lack of HI emission in observations toward Cygnus A.
They estimated the spin temperature of the WNM to be in the range of
3000 to 8000 K in different velocity ranges toward Cygnus A.
No absorption in the WNM was detected in any of the other observations 
mentioned above leading to upper limits on its optical depth and hence lower limits to the
spin temperature. The highest of these lower limits to the spin 
temperature is 10$^{4}$ K (Kulkarni et al. 1985).

Of all the phases of the interstellar medium, the WNM remains the least 
understood. Although HI 21 cm-line emission from the WNM is detected easily
in most directions, an accurate value for the temperature 
of the WNM is lacking. The spin temperature of the WNM is an important 
parameter with 
implications for models of the interstellar medium. The 
most detailed analysis of the equilibrium state of neutral hydrogen 
in the interstellar medium is the work of Wolfire et al. (1995).
The primary heating mechanism is photo-electric 
emission from dust grains, while cooling is dominated by fine
structure lines from heavy elements in colder regions, and by hydrogen 
recombination lines in warmer regions. 
Recently, we detected HI 21 cm-line absorption in the WNM using 
the Westerbork Synthesis Radio Telescope (WSRT). The absorption was detected 
toward Cygnus A at LSR velocities
of --40 km s$^{-1}$ and --70 km s$^{-1}$ (Figs. 3 and 4, and Table 1 in
Carilli, Dwarakanath, \& Goss 1998). These two velocity ranges were
previously identified as being relatively free of cold absorbing clouds.
The measured optical depth for the WNM along the line of sight to 
Cygnus A is (8.9 $\pm$ 1.9)$\times$ 10$^{-4}$ at --70 km s$^{-1}$ and
(8.5 $\pm$ 2.0)$\times$ 10$^{-4}$ at --40 km s$^{-1}$, with corresponding
spin temperatures of 6000 $\pm$ 1700 and 4800 $\pm$ 1600 K, respectively.
The volume filling factor for the WNM appears to be fairly high 
and is $\sim$ 0.4 (Carilli, Dwarakanath, \& Goss 1998).
The measured value of the spin temperature toward
Cygnus A is consistent with the range of kinetic temperature (5500 --
8700 K) for the WNM in the model proposed by Wolfire et al (1995). 
While the current data toward
Cygnus A are consistent with this picture, values of
the spin temperature of the WNM in more directions are needed to provide a 
constrained physical model of the interstellar medium.

Mebold et al. (1981) have carried out a survey
in Galactic HI absorption and emission toward a large number of extragalactic
radio  sources. From this list we have chosen four sources which
 are strong enough ($>$ 10 Jy at 20 cm) to
detect HI absorption (optical depth $\sim$ 0.001) in an 
integration time of $\sim$ 12 hours. 
The HI emission profile toward each of these sources shows a few  
 narrow lines plus a broad line component.  The existing HI absorption spectrum 
toward each of these sources, however, 
has only narrow components corresponding to the narrow 
emission lines(Mebold et al. 1982).
These absorption components (optical depth $>$ 0.05) are due to the cold gas 
(T$_{s}$ $\sim$ 80 K) in the Galaxy.
The motivation of the present observations is to detect the HI absorption
corresponding to the broad line in the
HI emission profile arising from the intercloud medium (Warm Neutral Medium).
The four sources chosen
for the present observations are at b $>$ 10$^{o}$.
These higher latitudes are relatively free of the Cold Neutral Medium, and because
of this, HI absorption
measurements toward high-latitude sources are less confused by the plethora of narrow
absorption lines  routinely detected in the Galactic plane. 

The observations are discussed in Section 2. The data analyses and the results obtained
are given in Section 3. The implications for the determinations of the temperature and 
the density of the Warm Neutral
Medium based on the current observation are discussed in Section 4.1. A by-product of the
current observations is the detection of HI 21 cm-line absorption from high velocity
clouds in the Outer Arm of the
Galaxy. A discussion on this can be found in Section 4.2.

\section{OBSERVATIONS}

The HI 21 cm-line absorption measurements were 
carried out toward 3C147, 3C273, 3C295 and 3C380 using the 
Westerbork Synthesis Radio Telescope (WSRT) in 1998 and 2000. 
Some details of the observations are given in Table 1.
The observations used a bandwidth of 1.25 MHz 
centered at the frequency of the HI 21 cm-line and 256 spectral channels. 
The resulting velocity resolution is $\sim$ 2.1 km s$^{-1}$ after Hanning smoothing.
Absolute flux density calibration was obtained by observing 3C286 and 3C48. 
Since the sources of interest were bright and unresolved at the WSRT 
resolutions ($\sim$ 15$''$), amplitude and phase calibrations of the antennas were carried 
out through self-calibration.  The bandpass calibration
was obtained through frequency switching on the source itself. 
For this purpose, each `on-line' scan of 25 minutes duration was sandwiched
between two `off-line' scans each of 25 minutes duration. These `off-line'
scans were offset in frequency from the `on-line' scan by $\pm$ 1.25 MHz 
respectively. The `off-line' scans are unaffected by contamination due to Galactic HI 
and provide an accurate bandpass. 

The data were analyzed using the Astronomical Image Processing System (AIPS) developed
by the National Radio Astronomy Observatory. For each source, a continuum data base was
constructed by averaging the line-free channels. The amplitude and phase solutions 
of antennas obtained through normal and self-calibration of this data set were
transferred to the spectral line data. The calibrated and bandpass corrected spectral line
data of the source
were used in further analysis. The continuum flux density in the source visibilities
was removed by making a linear fit to the visibilities in the line-free channels as 
a function of channel and subtracting it from the source visibilities in all the 
spectral channels. These continuum-subtracted visibilities were used to make spectral 
cubes of HI absorption images. The spectral cubes were CLEANed. The rms per channel
was typically $\sim$ 5 mJy/beam.  In order to avoid any possible contamination
due to Galactic HI emission, visibilities in the short spacings (the inner $\sim$ 0.5 km)
were not used in making the HI absorption images. 

Since the aim of the current observations is to detect weak (optical depth $\sim$10$^{-3}$),
and wide (width $\sim$ 50 km s$^{-1}$) lines, the most important requirement 
is the stability of the bandpass. This stability can be estimated by comparing the two
`off-line' bandpasses from the same source in a given observing run. Such 
a comparison for 3C147, 3C273 and 3C295 indicated that the two `off-line' 
bandpasses on a given source are consistent with each other to within 
2$\times$10$^{-4}$ (1$\sigma$). The calibrated on-line data of these sources is not
expected to have any spurious features beyond this level.
 This stability is greater than, or comparable to 
the rms noise in the optical depth images toward these three sources (Table 1).
In the case of 3C380, the two `off-line' bandpasses differed at 
the level of 6$\times$10$^{-4}$ (1$\sigma$). In addition, in the case of 3C380,
many antennas showed variations across the passband with periods
of $\sim$ 0.5 MHz ($\sim$ 100 km s$^{-1}$).
Due to the limited dynamic range achieved, the 3C380 data
were not considered for wide-line analysis. However, these data were suitable for
an investigation of narrow HI absorption lines. 

\section{RESULTS}

The HI absorption spectra toward the four sources in Table 1 are displayed in the lower
panels in Figs. 1--4 respectively. The corresponding HI emission profiles from the
Leiden Dwingeloo Sky Survey (Hartmann, \& Burton 1997) are in the upper panels of these
figures. A Gaussian decomposition of these profiles was carried out using the 
Groningen Image Processing System (GIPSY). Each optical depth spectrum was Hanning
smoothed and simultaneously fit with the
minimum number of Gaussians until the reduced chi-squared value was $\leq$ 1.
In each case, the model spectrum was then subtracted 
from the observed spectrum
to obtain the residual spectrum. The best-fit model spectrum shows a featureless
residual spectrum. The residual spectra so obtained toward the sources 3C147, 3C273, 
and 3C295 are shown in Fig. 7. Similar analyses were also carried out on the corresponding
HI emission spectra. The parameters of the best-fit Gaussians from these analyses
 are summarized in Table 2. In many cases, the central velocities of HI absorption
and HI emission components agree within errors. In such cases, we assume that the absorption
and emission arise from the same HI. An estimate of the column density,
the spin temperature, and the width expected due to thermal broadening is also
given for these components. If the central velocities do not agree within the
quoted errors, either the absorption
or the emission column is left blank, indicating that they arise from different HI. 

In Figs. 5 and 6, the HI absorption spectra toward 3C147 and 3C380 are shown along with 
the corresponding HI emission spectra respectively. Toward the direction of 3C147
note the weak HI absorption (Fig. 5b) over the velocity range
--70 km s$^{-1} <$ V$_{lsr} <$ --30 km s$^{-1}$  and a corresponding HI emission
(Fig. 5a).  The weak HI absorption detected toward 3C147 over the velocity range 
--70 km s$^{-1} <$ V$_{lsr} <$ --30 km s$^{-1}$ (Fig. 5b)
has a 10$\sigma$ significance compared to the bandpass stability observed in the 3C147
observations. This HI absorption can be combined with the corresponding
HI emission (Fig. 5a) to estimate the spin temperature of the HI gas responsible for them.  
The HI emission spectrum in
T$_{b}$ toward 3C147 (Fig. 5a) is divided by the optical depth
values (Fig. 5b) to obtain spin
temperature as a function of V$_{lsr}$. Over the velocity range 
--70 km s$^{-1} <$ V$_{lsr} <$ --30 km s$^{-1}$ where the weak absorption is detected, 
the spin temperature has an average value of 3000$\pm$500 K (1$\sigma$ error). 
The high spin temperature and the wide range of velocities over which HI absorption is
detected indicates that this HI gas  is due to the Warm Neutral Medium of the Galaxy. 
Note that the HI
absorption toward 3C380 (Fig. 6b) also shows a similar $'$dip$'$ over 
the velocity range --50 km s$^{-1} <$ V$_{lsr} <$ --20 km s$^{-1}$. This feature has less than
a 3$\sigma$ significance compared to the bandpass stability in the 3C380 observations.

In Fig. 7, the residual HI absorption spectra obtained after subtracting the best-fit models 
(Table 2) from the observed spectra toward 3C147, 3C273 and 3C295 are 
shown in top to bottom panels, respectively.  In the case of 3C147 (Fig. 7a),
the velocity range over which the simultaneous fit was carried out was restricted to
--110 km s$^{-1} <$ V$_{lsr} <$ +100 km s$^{-1}$ in order not to exceed the maximum
number of Gaussians that GIPSY can fit simultaneously. The spectrum shown for 3C147
(Fig. 7a) is restricted to this velocity range. A separate simultaneous fit
was carried out with two Gaussians over the velocity range --140 km s$^{-1} <$ 
V$_{lsr} <$ --90 km s$^{-1}$ to account for the two high velocity narrow 
absorption lines detected toward 3C147 (Fig. 5b and Table 2). 
There is a clear separation between these
two velocity ranges of the spectrum with no spectral features in the overlapping region 
--110 km s$^{-1} <$ V$_{lsr}<$ -90 km s$^{-1}$ (Fig. 5b) 
enabling independent Gaussian analyses in the two velocity ranges.
In Fig. 7a two spectra corresponding to 3C147 are shown. The solid line
indicates the residual absorption spectrum obtained after subtracting from the
observed spectrum the model containing only the seven narrow
HI absorption lines in the velocity range 
--110 km s$^{-1} <$ V$_{lsr} <$ +100 km s$^{-1}$ (Table 2).
The broken line indicates the residual spectrum 
after including the wide HI absorption line
(V$_{lsr} =$ --29$\pm$4 km s$^{-1}$ and $\delta$V$_{1/2}$ = 53$\pm$6 km s$^{-1}$)
in addition to the seven narrow lines in the model spectrum. The reduced chi-squared values 
of the residual spectra are 16 (solid line) and 0.6 (broken line) respectively. 
These two residual spectra and the corresponding reduced chi-squared values clearly 
indicate the existence of a wide HI absorption line toward 3C147. In addition, a 
corresponding wide ($\delta$V$_{1/2}$ = 34$\pm$1.1 km s$^{-1}$) HI emission component at a 
velocity V$_{lsr} =$ --35.5$\pm$0.3 km s$^{-1}$ is observed in the LDSS ( Fig. 5a and Table 2). 
The central velocities of these absorption and
emission lines agree to within $\sim$1$\sigma$. However, their widths  are discrepant at
the 3$\sigma$ level. 
We identify these two emission and absorption lines as arising from the same HI gas. 
The estimated spin temperature of this component is 3600 $\pm$ 360 K, in agreement
with the direct estimate of spin temperature made in the previous paragraph by 
dividing the HI emission brightness 
temperature by the corresponding optical depth as a function of V$_{lsr}$. 


The residual spectra shown
in Figs. 7b,c corresponding to the HI absorption toward 
3C273 and 3C295 do not indicate wide HI absorption, although the
HI emission spectra toward these sources do show wide lines (Table 2) .
In these two cases upper limits to the optical depths of wide absorption
lines can be obtained. Based on these upper limits and the HI column densities 
of the corresponding emission
lines, lower limits to the spin temperatures of these wide emission 
lines are estimated (Table 2). For example, toward 3C273, a lower limit (3$\sigma$) of
2600 K is obtained for the emission line at V$_{lsr}$ = -- 13.2 km s$^{-1}$ with a 
width of $\delta$V$_{1/2}$ = 24.7$\pm$1.4 km s$^{-1}$. 

The HI absorption profile toward
3C147 (Fig. 5b) also shows two narrow, weak HI absorption lines at 
V$_{lsr}$ = --117.3 and --124.5 km s$^{-1}$ due to HI in  the Outer Arm of the Galaxy. 
The HI emission profile toward 3C147 (Fig. 5a) has
a line at V$_{lsr}$ = --120.4 km s$^{-1}$ 
with a full width at half maximum of 17 km s$^{-1}$ (Table 2). 
It appears that this HI emission and the 
two HI absorption lines may be associated. However, since there is no exact correspondence
of the central velocities and the widths of the emission and absorption lines, we can
only estimate limits on the spin temperatures. Assuming that the widths
of these two absorption lines are due to thermal broadening alone, upper limits to their spin
temperatures of 1400 K and 270 K are obtained, respectively. 
The absence of HI absorption corresponding to the
HI emission at V$_{lsr}$ = --120.4 km s$^{-1}$ leads to a lower limit of 1780 K to its
 spin temperature.
Similarly, the HI absorption spectrum toward 3C380 (Fig. 6b)
has a discrete narrow line at V$_{lsr} = -$113.7 km s$^{-1}$ with an
HI emission line in the LDSS spectrum at 
V$_{lsr}$ = --109.7 km s$^{-1}$ (Fig. 6a).  However, HI emission line 
width is more than ten times the HI absorption line width (Table 2).
The lower limit to the spin temperature of this wide emission line with no corresponding
HI absorption is 800 K. 
An upper limit to the spin temperature of 210 K can be estimated for
the HI absorption  at V$_{lsr} = -$113.7 km s$^{-1}$, if the width is due entirely 
to thermal broadening. 

\section{DISCUSSION}

\subsection{Temperature and Density}

The spin temperature of the HI gas is influenced by the intensity of radiation at the frequency
of the 21 cm-line, collisions with electrons, protons, and hydrogen atoms, and the intensity of 
the Ly$\alpha$ radiation field (Field 1958, Liszt 2001). In the context
of the Warm Neutral Medium, the most important factor influencing the spin temperature of the HI gas
is the Ly$\alpha$ radiation field. The HI spin temperature is expected  
to be close to the kinetic temperature of the gas through the process of
 resonant scattering of ambient Ly$\alpha$
photons (Field 1958). The efficacy of this process depends on the ambient density of 
Ly$\alpha$ photons. There are no secure measurements of the diffuse Galactic Ly$\alpha$ photon
density; an upper limit of 2.5$\times$10$^{-6}$ cm$^{-3}$ in the 
solar neighborhood  has been estimated (Holberg 1986).  The areal production rate
of ionizing photons from early-type stars in the solar neighborhood is estimated to be 
3.7$\times$10$^{7}$ photons cm$^{-2}$ s$^{-1}$ (Vacca, Garmany, \& Shull 1996). Assuming this
to be comparable to the production rate of Ly$_{\alpha}$ photons leads to a Galactic 
Ly$\alpha$ flux from early-type stars.  
Liszt (2001) discusses the relation between the spin temperature and the
kinetic temperature as a function of the number density of H-nuclei, the gas pressure
of the interstellar medium, and the fraction of Galactic Ly$\alpha$ flux 
from early-type stars that permeates the diffuse gas. 

If the fraction of Ly$\alpha$ 
flux from early-type stars that permeates the warm gas is $<$0.0001, 
the spin temperature is significantly lower than the kinetic temperature of the HI gas.
The spin temperature of 3600 K estimated in the current observations
implies a kinetic temperature of 7000 K and a density of 0.4 cm$^{-3}$ 
in the two-phase model proposed by Wolfire et al (1995) (Fig. 5 in Liszt 2001).
On the other hand, if the fraction of Ly$\alpha$ flux from early-type stars
that permeates the warm gas is large ($>$0.01), the spin temperature is within
10\% of the kinetic temperature. However, HI gas at such a low kinetic temperature as 
3600 K does not have a stable phase and is outside the allowed range of values in the model
proposed by Wolfire et al (1995) (Fig. 5 in Liszt 2001). Either we are observing an unstable 
phase of HI, or very small, if any, fraction of Ly$\alpha$ radiation from 
early-type stars permeates the diffuse intercloud medium.

The warm HI gas at V$_{lsr} \sim$ --30 km s$^{-1}$ is associated with the Perseus Arm of the Galaxy 
at a distance
of 2.5 kpc (l$\sim$ 160$^o$, b$\sim$ 10$^o$). The mean z 
of the HI gas is thus $\sim$ 460 pc. 
At this height the electron density is $<$ 0.02 cm$^{-3}$ 
(Taylor \& Cordes 1993), and  
combined with the current estimate of n$_{HI}$ implies that the ionization fraction at this
height is $<$ 0.05. As discussed earlier in this section, the
kinetic temperature of the WNM in the direction of 3C147 can be in the range 3600 - 7000 K.
The full width at half maximum of the HI 21cm emission line from the WNM
 due to thermal broadening alone is 13 - 18 km s$^{-1}$. The observed HI emission line
width from the WNM in the direction of 3C147 is 34 km s$^{-1}$. If the observed excess width 
is attributed to turbulence in the WNM, the implied full width at half maximum of the 
turbulence broadening is 31 - 29 km s$^{-1}$. 

\subsection{The Outer Arm and High Velocity HI Gas}


The Outer Arm is the structure observed in HI in the range 
50$^o < l < $ 195$^o$, --5$^o < b < $ 35$^o$, and
--175 $<$ V$_{lsr}$ $<$ --60 km s$^{-1}$ (Wakker, \& van Woerden 1991). 
The Outer Arm appears to lie between 15 and 20 kpc from the Galactic
Center in this longitude range (Kulkarni, Blitz, \& Heiles 1982). 
At the latitudes of 3C147 and 3C380 the HI emission images from 
the LDSS show well-delineated structures in the longitude-velocity plane 
which can be associated with the Outer Arm. In addition, 
there are discrete HI emission features in these images at the longitudes
of 3C147 and 3C380. Corresponding to these features, the HI emission spectra 
toward 3C147 and 3C380
from LDSS (Figs. 5a and 6a) clearly show lines at 
V$_{lsr}$ = --120.4 and --109.7 km s$^{-1}$ respectively (Table 2).
The high velocity HI absorption lines detected in the current observations toward 
3C147 (V$_{lsr}$ = --117.3 and --124.5 km s$^{-1}$) and 3C380 
(V$_{lsr}$ = --113.7 km s$^{-1}$) (Figs. 5b and 6b) are associated with the respective HI
emission features arising in the Outer Arm.
As mentioned in Section 3, in both cases (i.e., 3C147 and 3C380),
the high velocity clouds have at least two temperature components -- one with a 
spin temperature higher than,  
or comparable to 1000 K, and the second with a spin temperature lower than 200 K. 
The higher temperature gas, 
presumably wide-spread, is observed in HI emission. The narrow line HI absorption
may arise from smaller and colder regions (`clumps') embedded in this bigger structure. 
There are probably many such `clumps' in the high velocity HI emission features 
since HI absorption was detected toward two out of the four extragalactic sources observed. 

The heliocentric distances to these high velocity HI clouds toward 3C147 and 3C380 are 10 kpc
and 18 kpc respectively, assuming a mean galactocentric distance of 18 kpc to the Outer Arm
(Kulkarni, Blitz, \& Heiles 1982, Blitz, Fich, \& Stark 1982).
At these distances, the heights of these high velocity clouds above the Galactic
plane are 1.8 kpc and 8 kpc respectively.  
Very little is known about the spin temperature of HI in the Outer Arm
due to paucity of HI absorption measurements. Payne, Salpeter, \& Terzian (1980) 
reported an HI absorption measurement in the Outer Arm toward 4C33.48 (l=66.4$^o$, b=8.4$^o$)
 at V$_{lsr}$ = --128 km s$^{-1}$ with $\tau$ = 0.011 and T$_{s}$ = 69 K. 
More recently, Akeson, \& Blitz (1999) detected HI absorption in the Outer Arm
 toward the
extragalactic sources 1901+319 and 2005+403. 
They report $\tau$ = 0.012 and 0.031 and T$_{s}$ = 38 and 147 K
for these two lines of sight respectively. The widths of these HI absorption lines and the corresponding
HI emission lines are between 5 and 12 km s$^{-1}$. Although the current HI absorption in the
Outer Arm seems to imply the existence of similar spin temperature components, 
there are two distinct 
differences compared to the earlier two measurements : (1) the optical depths observed in the
current observations are about a factor of ten smaller than the previous measurements,
and (2) the current measurements indicate the existence of multiple 
temperature components in the Outer Arm HI. The existence of multiple temperature
components in other parts of the Galaxy has been well-established based on
earlier HI observations (Mebold et al 1982, Stark et al 1994). 
It appears that the less explored Outer Arm is not too different in this respect. 
However, the Outer Arm is different from the inner Galaxy in at least two aspects - 
(1) the column density of atomic Hydrogen can be less than or equal to 10$^{19}$ cm$^{-2}$,
and (2) the flux of diffuse Ly$_{\alpha}$ radiation can be significantly lesser. 
These two factors play an important role in estimating the allowed range of spin 
temperatures for an allowed range of kinetic temperatures in the two-phase 
equilibrium models of Wolfire et al (1995). It will be interesting to compare these
estimates of spin temperatures with those obtained from observations toward many 
lines of sight in the Outer Arm. 

\section{CONCLUSIONS}

We present observations of the Galactic HI 21 cm-line absorption toward four
extragalactic sources using the
WSRT. Towards one of the sources we have detected HI 21 cm-line absorption 
in the Warm Neutral Medium associated
with the Perseus Arm of the Galaxy. The observed optical depth is 
(1.9$\pm$0.2)$\times$10$^{-3}$
leading to a spin temperature estimate of 3600$\pm$360 K. Depending on the intensity
of the diffuse Ly$\alpha$ radiation field the kinetic temperature of the WNM is 
in the range 3600 - 7000 K.

We have also detected HI 21 cm-line absorption from high velocity clouds in the
Outer Arm of the Galaxy. We find multiple temperature components (T$_{s} <$ 200 K
and $>$ 1000K) in these high velocity clouds similar to that observed in other
parts of the Galaxy.

\acknowledgements We thank the referee for a thorough reading of an earlier version
of this paper and giving detailed comments to improve the clarity and readability
of the paper. 
The National Radio Astronomy Observatory (NRAO) is a facility of the National 
Science Foundation, operated under cooperative agreement by Associated
Universities, Inc..
The Westerbork Synthesis Radio Telescope  is operated by the Netherlands Foundation
for Research in Astronomy with financial support from the Netherlands Organization
for Scientific Research (NWO). We thank Rene Vermeulen for scheduling and 
carrying out these observations.

\clearpage

\begin{figure}
\figurenum{1}
\epsscale{0.7}
\plotone{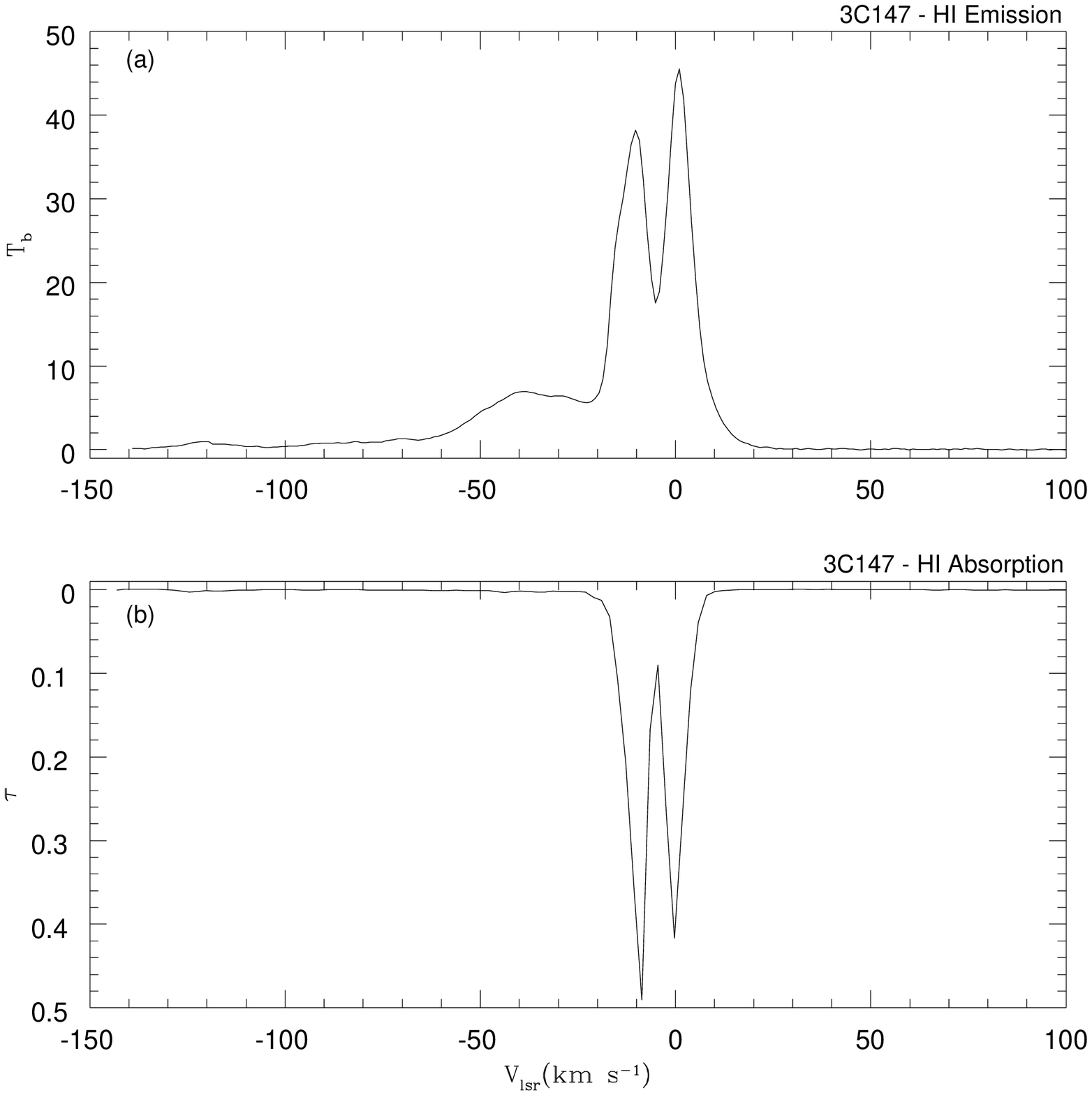}
\caption{
(a) HI emission at l=161.5$^o$, b=10.5$^o$ from the Leiden Dwingeloo Sky Survey
(LDSS) (Hartmann, \& Burton 1997). 
The LDSS has a resolution of $\sim$ 35$'$, an rms of $\sim$ 0.07 K and a velocity resolution
of $\sim$ 1 km s$^{-1}$. 
(b) HI absorption toward 3C147 from the  current observations
using the Westerbork Synthesis Radio Telescope. The WSRT observations have a 
synthesized beam of
16.6$'' \times$12.1$''$ (P.A.=--2.4$^o$), an rms of 1.5 $\times$ 10$^{-4}$ in $\tau$ and
a velocity resolution of $\sim$ 2.1 km s$^{-1}$. The x-axis is velocity in the 
Local Standard of Rest.} 
\end{figure}

\begin{figure}
\figurenum{2}
\epsscale{0.7}
\plotone{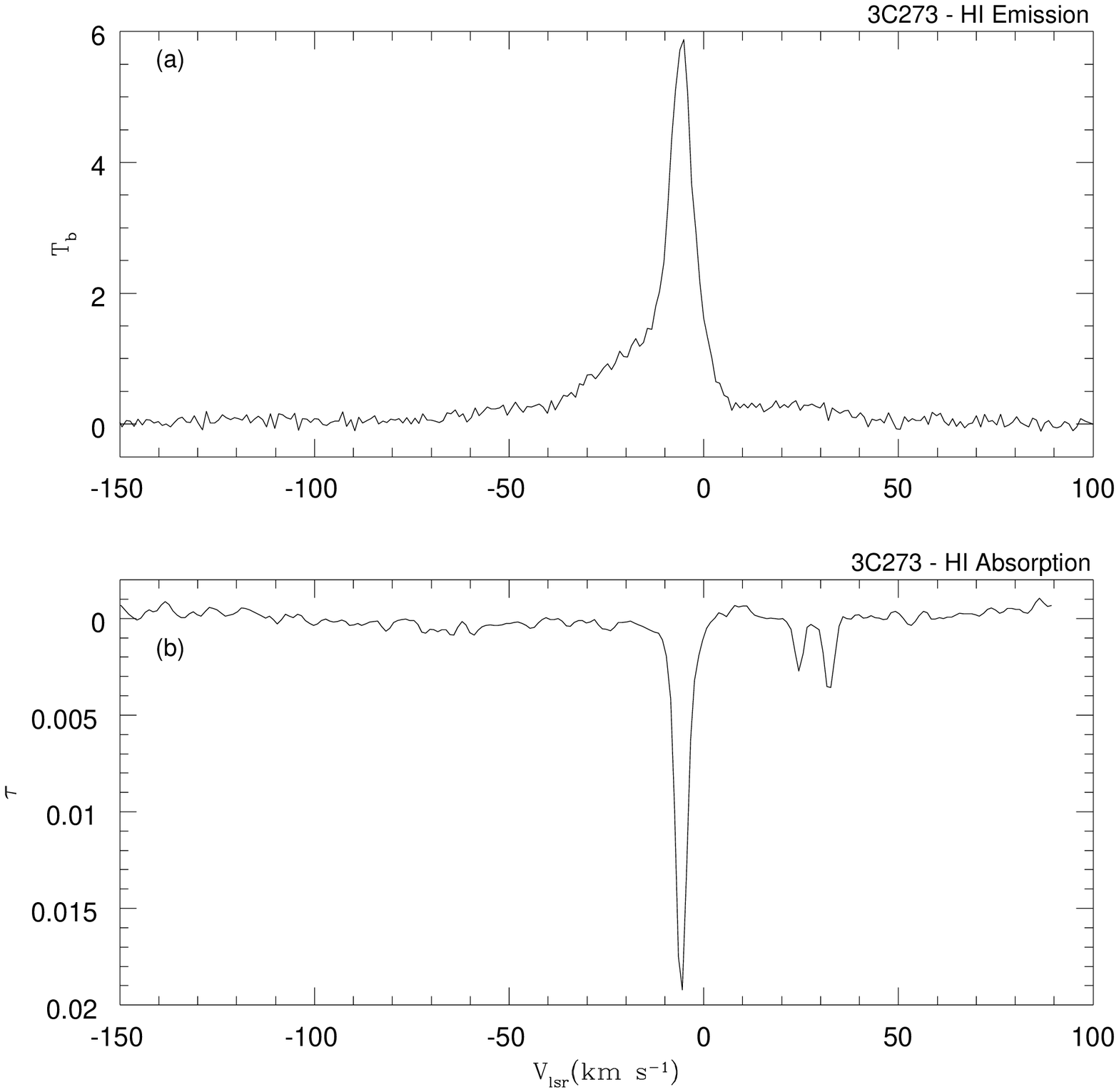}
\caption{
(a) HI emission at l=290.0$^o$, b=64.5$^o$ from LDSS.
(b) HI absorption toward 3C273 from the WSRT. 
The WSRT observations have a synthesized beam of 511$'' \times$11.7$''$ (P.A.=0.4$^o$),
an rms of 1.2$\times$10$^{-4}$ in $\tau$ and a velocity resolution of $\sim$ 2.1 km s$^{-1}$.} 
\end{figure}

\begin{figure}
\figurenum{3}
\epsscale{0.7}
\plotone{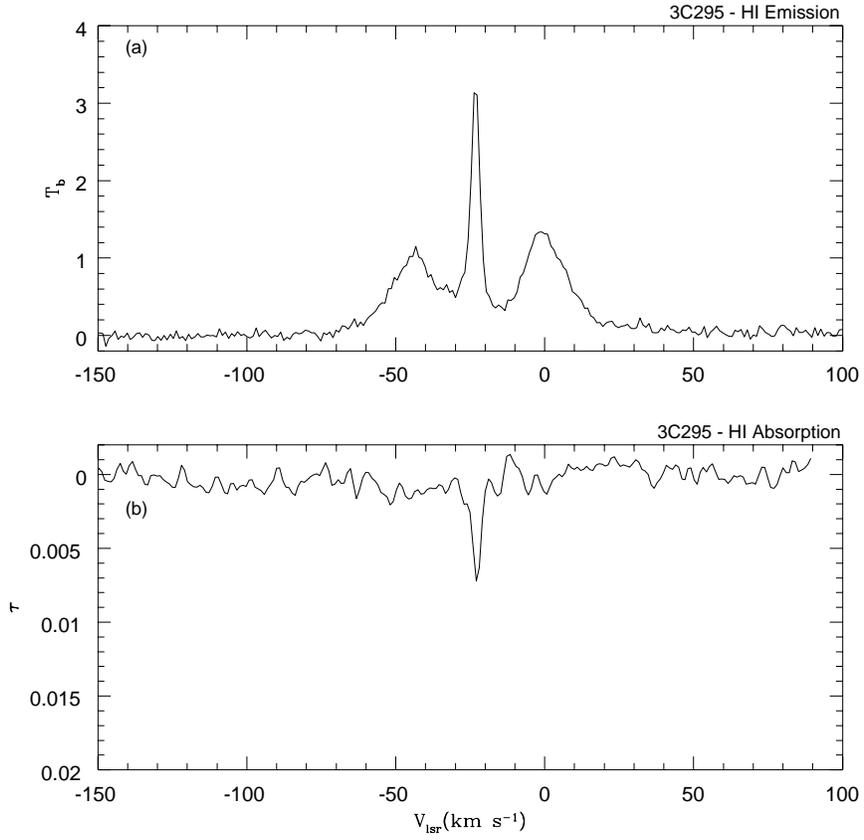}
\caption{
(a) HI emission at l=97.5$^o$, b=61.0$^o$ from LDSS.
(b) HI absorption toward 3C295 from the WSRT. 
The WSRT observations have a synthesized beam of 
 28.9$'' \times$14.4$''$ (P.A.=--75.0$^o$), an rms of 3.6$\times$10$^{-4}$ in $\tau$ and
a velocity resolution of $\sim$ 2.1 km s$^{-1}$.}
\end{figure}

\begin{figure}
\figurenum{4}
\epsscale{0.7}
\plotone{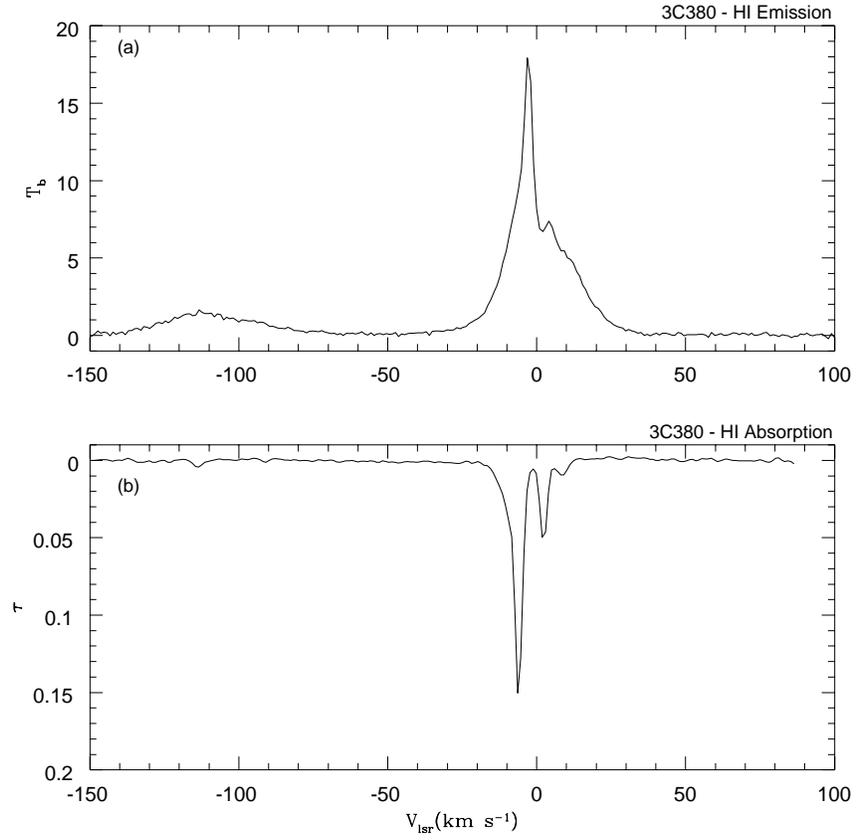}
\caption{
(a) HI emission at l=77.0$^o$, b=23.5$^o$ from LDSS.
(b) HI absorption toward 3C380 from the WSRT. 
The WSRT observations have a synthesized beam of 46.1$'' \times$19.7$''$ (P.A.=16.1$^o$),
an rms of 5.2$\times$10$^{-4}$ in $\tau$ and a velocity resolution of $\sim$ 2.1 km s$^{-1}$.}
\end{figure}

\begin{figure}
\figurenum{5}
\epsscale{0.7}
\plotone{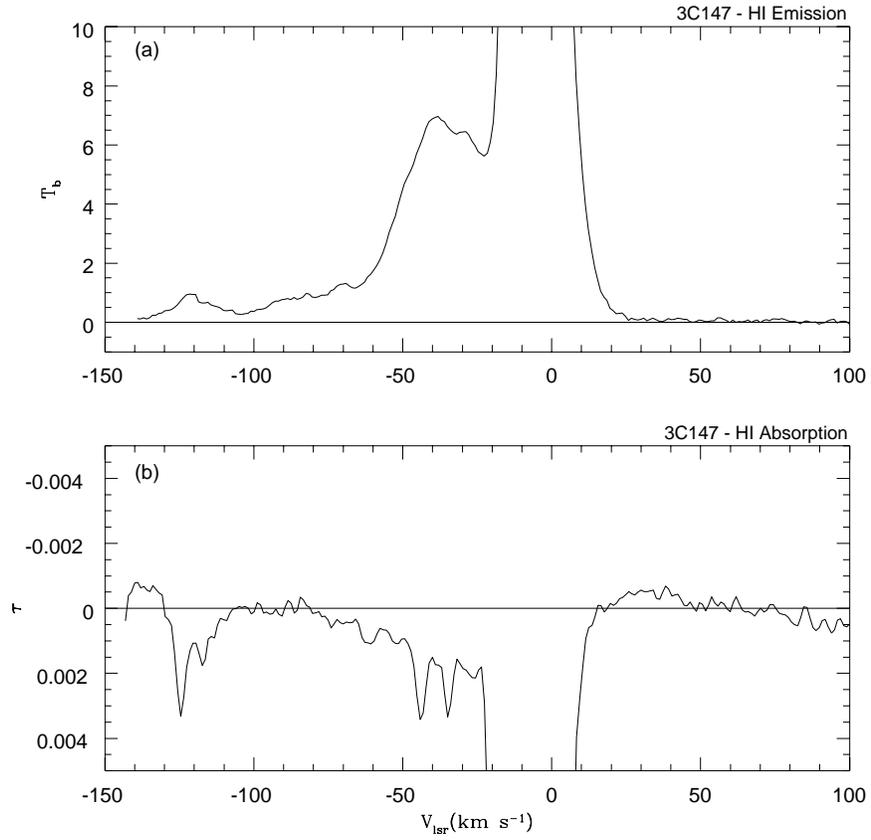}
\caption{ Same as Fig. 1 in an expanded y-scale. Note the weak HI absorption over
the velocity range --70 km s$^{-1} <$ V$_{lsr} <$ --30 km s$^{-1}$ in (b) and
a corresponding HI emission in (a). The implied spin temperature (T$_{b}$/$\tau$) over
this velocity range is 3000$\pm$500 K.}
\end{figure}

\begin{figure}
\figurenum{6}
\epsscale{0.7}
\plotone{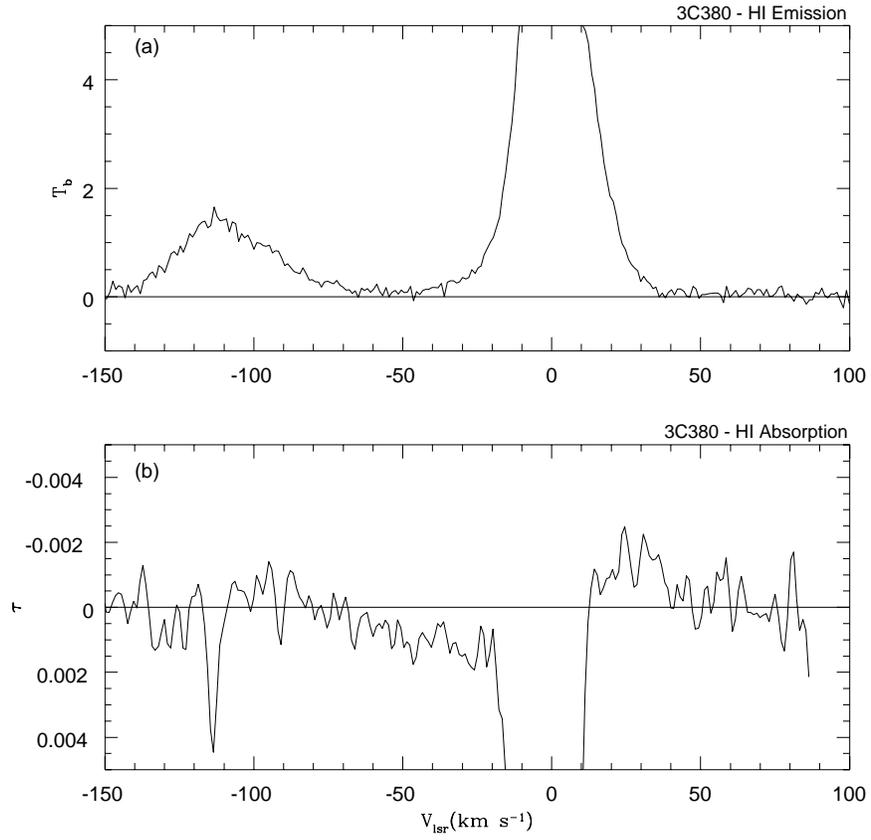}
\caption{ Same as Fig. 4 in an expanded y-scale. }
\end{figure}

\begin{figure}
\figurenum{7}
\epsscale{0.7}
\plotone{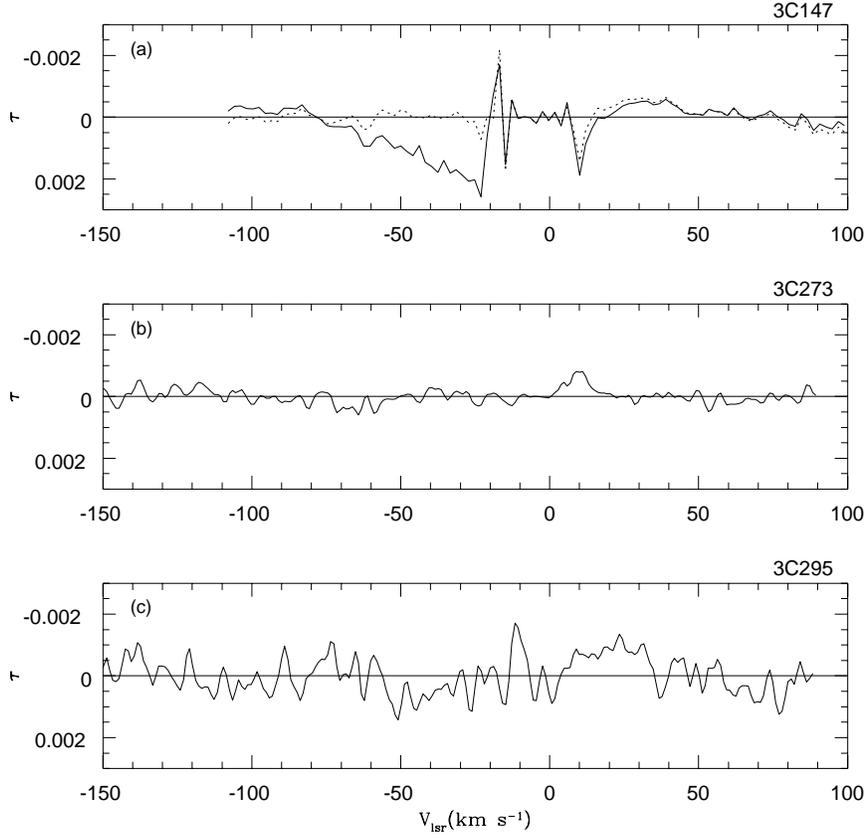}
\caption{Residual HI absorption spectra (data--model) toward 3C147, 3C273 and 3C295. 
The parameters of the best-fit HI absorption Gaussian components used in the models
are listed in Table 2. 
(a) The solid line shows the residual spectrum toward 3C147 when 
the model included only the 7 narrow HI absorption lines 
in the velocity range --110 km s$^{-1} <$ V$_{lsr} < $ 100 km s$^{-1}$.
The broken line is the residual spectrum 
when the model included the wide absorption line of width 53 km s$^{-1}$ at 
V$_{lsr}$ = --29 km s$^{-1}$. 
The sharp features at V$_{lsr} \sim$ --15 and +10 km s$^{-1}$
are artifacts left over from fitting multiple Gaussians to deep and narrow lines. 
(b) The residual spectrum toward 3C273. 
The amplitudes of the artifacts in (a) are a small percent of the peak optical depth
and would not be detected in this spectrum. 
(c) The residual spectrum toward 3C295.}

\end{figure}

\clearpage
 






\begin{deluxetable}{cccccccc}
\tabletypesize{\footnotesize}
\tablenum{1}
\tablecolumns{8}
\tablewidth{0pc}
\tablecaption{WSRT Observations}
\tablehead{
\colhead{Source}&\colhead{Galactic Coord.}&\colhead{S$_{1420}$} 
&\colhead{Synthesized Beam}&\colhead{P.A.}&\colhead{RMS}&\colhead{$\tau$} & \colhead{Time}
\\
\colhead{}&\colhead{(deg)}&\colhead{(Jy)}&\colhead{(FWHM in arcsec)} 
&\colhead{(deg)}&\colhead{(mJy beam$^{-1}$)}&\colhead{(RMS)} & \colhead{(hr)}
}
\startdata
3C147 & 161.7+10.3 & 21.3 & 16.6 $\times$ 12.1 & --2.4 & 3.3 & 1.5$\times$10$^{-4}$ & 2 $\times$ 15 \nl
3C273 & 290.0+64.4 & 26.9 & 511 $\times$ 11.7 & 0.4 & 3.2 & 1.2$\times$10$^{-4}$ & 1 $\times$ 12  \nl
3C295 & 97.5+60.8 & 21.3 & 28.9 $\times$ 14.4 & --75.0 & 7.7 & 3.6$\times$10$^{-4}$ & 2 $\times$ 10 \nl
3C380 & 77.2+23.5 & 12.3 & 46.1 $\times$ 19.7 & 16.1 & 6.4 & 5.2$\times$10$^{-4}$ & 1 $\times$ 12 \nl
\enddata
\tablecomments { All the observations had a velocity resolution of 2.1  km s$^{-1}$ and
128 channels after Hanning smoothing. The RMS quoted is per channel. 
The last column indicates the approximate
total time spent on each observation. The observations were carried out 
during December 1998, March 2000 and May 2000.
 } 
\end{deluxetable}







\begin{deluxetable}{rrrrrrrrrr}
\tabletypesize{\footnotesize}
\tablenum{2}
\tablecolumns{10}
\tablewidth{0pc}
\tablecaption{Best-fit Gaussian components}
\tablehead{\colhead{Source} &
\multicolumn{3}{c}{HI absorption} & \multicolumn{4}{c}{HI Emission}
&\colhead{} &\colhead{} \\
\colhead{} & \colhead{V$_{lsr}$}&\colhead{$\delta$V$_{1/2}$}&\colhead{$\tau\tablenotemark{a}$}
&\colhead{V$_{lsr}$} 
&\colhead{$\delta$V$_{1/2}$}&\colhead{T$_{b}$}&\colhead{N$_{HI}$}&\colhead{T$_{s}$}
&\colhead{$\delta$V$_{D}$}\\
\colhead{} & \colhead{(km/s)}&\colhead{(km/s)}&\colhead{}&\colhead{(km/s)} 
&\colhead{(km/s)}&\colhead{(K)}&\colhead{}&\colhead{(K)}
&\colhead{(km/s)}
}
\startdata
3C147 & +2.8(.5) & 5.0(.4) & .1(19) & +5.6(2) & 11.1(2) & 8.5(2)&1.8 & 85 & 2.0\nl
& --0.6(.1) & 4.6(.1) & .38(21) & +0.7(.04) & 6.5(.2) & 39.5(4.1)& 5.0 & 105 & 2.2 \nl
& --8.8(.01) & 2.9(.04) & .29(5) & & & &  & $<$185 & \nl
& --10.7(.06) & 7.3(.06) & .26(3) & --10.8(.04) & 9.6(.1) & 35.9(.2)& 6.7 & 140 & 2.5 \nl
& --20.2(.07) & 2.4(.8) & .009(.7) & & & & & $<$125 &  \nl
& --29(4) & 53(6) & .0019(.2) & --35.5(.3) & 34(1.1) & 6.9(.1)& 4.6 & 3600 & 12.8 \nl
& --34.7(.8) & 2.7(3.9) & .0013(1.5) & & & & & $<$160 & \nl
& --43.9(.6) & 3.5(1.2) & .0016(.5) & & & & & $<$270 & \nl
&        &    &        & --81.8(2.2) & 29(6) & 0.8(.1) & 0.5 & $>$1780 &  \nl 
& & & & --120.4(1.4) & 17(3.6) & 0.8(.1) & 0.3 & $>$1780 & \nl
& --117.3(.3) & 8.0(.9) & .0013(.1) &  & &  & & $<$1400 &  \nl
& --124.5(.1) & 3.5(.2) & .003(.1) &  &  &  &  & $<$270 & \nl 
\cline{1-10} \\
3C273 & +32.2(.08) & 3.0(.2) & .0038(.2) & & &  & & $<$200 &  \nl
& +24.5(.1) & 2.6(.3) & .0026(.2) & &  & & & $<$150 & \nl
& --5.4(.1) & 6.6(.6) & .0048(1.1) & &   & &    & $<$950 &      \nl
& --5.8(.03) & 3.0(.1) & .0147(1.1) & --5.7(.03) & 6.8(.1) & 4.6(.06) & 0.6 & 300 & 3.7 \nl
&       &     &        & --12.7(1.6) & 76.0(8) & 0.34(.05) & 0.5 & $>$900 &  \nl
&       &     &        & --13.2(.5) & 24.7(1.4) & 0.95(.06) & 0.5 & $>$2600 & \nl
\cline{1-10} \\
3C295 & & & &  --0.25(.2) & 19.1(.4) & 1.2(.02) & 0.4 & $>$1100 &  \nl
& --22.9(.1) & 3.5(.3) & .0066(.5) & --23.3(.03) & 3.9(.08) & 2.9(.05) & 0.2 & 440 & 4.5 \nl
&        &     &        & --41.9(.3) & 24.9(.7) & 0.9(.02) & 0.4 & $>$800 &  \nl
\cline{1-10} \\
3C380& +8.4(.1) & 4.2(.3) & .0098(.7) & &  & & & $<$385 & \nl
& +2.3(.02) & 2.8(.05) & .0526(.8) &  &  & & & $<$170  &  \nl
&     &     &        & +0.1(.1) & 26.6(.2) & 7.6(.1) & 3.9 & $>$4200 &  \nl
&     &     &        & --3.1(.02) & 3.5(.07) & 10.1(.2) & 0.7 & $<$270 &   \nl
& --6.0(.01)& 2.7(.03) & .1208(1.2) & & & & & $<$160 & \nl
& --8.3(.1) & 8.3(.1) & .0356(.1) &     &    & &      & $<$1510 &    \nl
&  &  &  &--109.7(.6) & 38.1(1.4) & 1.3(.04) & 1.0 & $>$800 &  \nl 
& --113.7(.3) & 3.1(.6) & .0043(.8) & & & & & $<$210 &  \nl 
\enddata
\tablenotetext{a} {The 1$\sigma$ errors obtained from fitting are given in parentheses in
units of 10$^{-3}$.}
\tablecomments { The HI emission components are best-fit to data from LDSS 
(Hartmann, \& Burton 1997). The HI absorption
data is from the current observations. N$_{HI}$ is in units of
10$^{20}$ cm$^{-2}$.  $\delta$V$_{1/2}$ is full 
width at half maximum. The lower limits
on T$_{s}$ are 3$\sigma$ limits based on the upper limits on $\tau$. The upper limits on T$_{s}$ 
correspond to the values if the line widths are entirely due to thermal broadening. 
In those cases where absorption and emission were detected
from the same gas, the width due to pure thermal broadening is given by
$\delta$V$_{D}$. The numbers in parentheses are 1$\sigma$ formal errors from the fitting
procedure. } 
\end{deluxetable}

\end{document}